\documentclass[10pt,twocolumn,reprint,amsmath,amssymb,aps,prl,showpacs,longbibliography,superscriptaddress,floatfix]{revtex4-1}
\usepackage[utf8]{inputenc}
\usepackage{amsmath}
\usepackage{amssymb}
\usepackage{graphicx}
\usepackage{xcolor}

\makeatletter
\usepackage{amsfonts}
\usepackage{graphicx}
\usepackage{graphics}

\makeatother

\begin{document}
\title{Motion-Induced Polaron Resonance beyond the Orthogonality Catastrophe}
\author{Jia Wang}
\affiliation{Centre for Quantum Technology Theory, Swinburne University of Technology,
Melbourne 3122, Australia}
\author{Xia-Ji Liu}
\affiliation{Centre for Quantum Technology Theory, Swinburne University of Technology,
Melbourne 3122, Australia}
\author{Hui Hu}
\affiliation{Centre for Quantum Technology Theory, Swinburne University of Technology,
Melbourne 3122, Australia}
\date{\today}
\begin{abstract}
  We investigate the spectroscopic response of a repulsively interacting, infinitely heavy impurity moving at constant velocity $v$ through a one-dimensional ideal Fermi gas. Using an exact functional-determinant approach, we calculate the Ramsey response and spectral function, fully accounting for many-body particle-hole excitations. We uncover a strongly nonmonotonic dependence on impurity velocity: decoherence is enhanced around the Fermi velocity, whereas sufficiently rapid motion restores a narrow, long-lived polaron resonance. A long-time wave-packet analysis yields analytical expressions for the polaron energy and decay rate over the full velocity range and quantitatively captures the numerical results. At high velocity, the polaron energy approaches the mean-field value, the decay rate vanishes as $1/v$, and the spectral weight approaches unity. Thus, rapid prescribed motion provides a dynamical route to suppressing the orthogonality catastrophe and restoring polaron coherence even for an infinitely heavy impurity, where recoil is entirely absent.
\end{abstract}
\maketitle

Fermi polaron, formed by an impurity dressed by excitations of a surrounding Fermi sea, is a paradigmatic quasiparticle in quantum many-body physics \cite{Chevy2006PRA}. Its properties reflect the interplay between impurity-fermion interactions and the many-body excitations they generate. Ultracold atomic gases provide a versatile platform for studying these effects, as interactions and impurity states can be precisely controlled \cite{Schirotzek2009PRL,Zhang2012PRL,Grimm2012Nature,Kohl2012Nature,Demler2016Science,Hu2016PRL,Jorgensen2016PRL,Scazza2017PRL,Yan2019PRL,Zwierlein2020Science,Sagi2020PRX,Lobo2006PRL,Combescot2007PRL,Punk2009PRA,Cui2010PRA,Mathy2011PRL,Schmidt2012PRA,Rath2013PRA,Shashi2014PRA,Li2014PRA,Kroiss2015PRL,Levinsen2015PRL,HuHui2016PRA,Goulko2016PRA,HuHui2018PRA,PenaArdila2019PRA,Mulkerin2019AnnPhys,Jia2019PRL,Isaule2021PRA,Pessoa2021PRA,Seetharam2021PRL}. In particular, impurity spectroscopy offers complementary probes of polaron formation: Ramsey interferometry measures the coherence between interacting and noninteracting impurity states, while radio-frequency (RF) spectroscopy resolves the response in frequency \cite{Demler2012PRX}. These techniques offer a route to addressing a fundamental question of how well-defined quasiparticles emerge despite the possibility of creating arbitrarily many particle-hole excitations. Though Chevy variational ansatz, which retains at most one particle-hole excitation, has proved remarkably successful in describing polaron energies and quasiparticle properties, the microscopic origin of Fermi polaron amid many particle-hole excitations remains elusive.

In this context, a contrasting limit arises for an infinitely heavy impurity immersed in an ideal Fermi gas \cite{Demler2012PRX,Schmidt2018Review}. The impurity acts as a scattering potential, whose sudden introduction can excite arbitrarily many low-energy particle-hole pairs. The resulting Anderson orthogonality catastrophe (OC) causes the overlap between the initial and final many-body states to vanish in the thermodynamic limit, eliminating the quasiparticle residue and producing a power-law singularity in the spectral response \cite{Anderson1967PRL}. The fate of this singularity depends sensitively on the mechanisms that regulate low-energy excitations. For a mobile impurity in three dimensions, finite-mass recoil can regularize the low-energy particle-hole excitations responsible for the OC, allowing a finite quasiparticle residue under appropriate conditions \cite{SchmidtChen2025PRL}. In one dimension, however, OC physics can persist even for a mobile impurity, producing singular spectral features rather than a conventional quasiparticle peak \cite{Castella1996PRB,Castella1993PRB, VanHoucke2026PRA, Guiliano2026PRA}. Related studies of moving impurities have examined momentum-resolved polaron spectra \cite{Parish2021PRA, Massignan2026arXiv} and the nonequilibrium dynamics of momentum injection, including quantum flutter in one-dimensional quantum gases \cite{Demler2012NP,Knap2014PRL}. Another route to suppressing the OC is to introduce a pairing gap in the surrounding Fermi superfluid, which can restore coherent polaron features \cite{Jia2022PRL,Jia2022PRA,Jia2022Review}. These results highlight the central role of low-energy excitations in determining the emergence of coherent polaron quasiparticles.

\begin{figure}[t]
  \centering
  \includegraphics[width=0.95 \columnwidth]{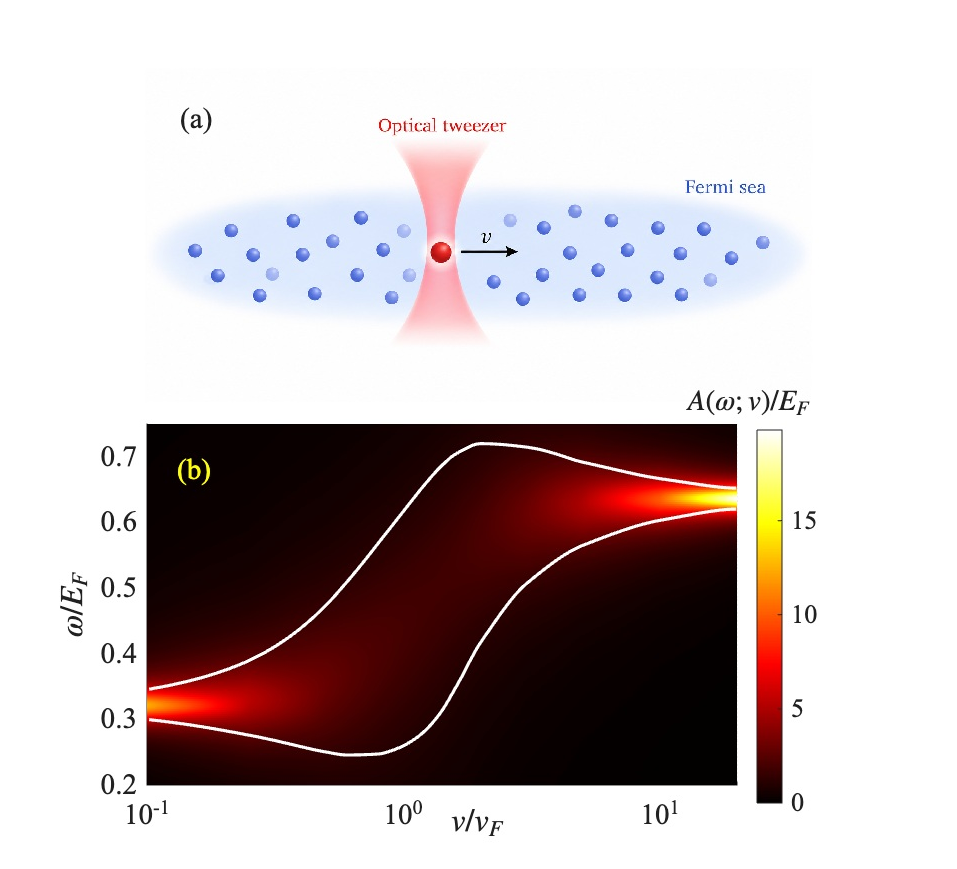}
  \caption{(a) Schematic of an impurity transported through a Fermi sea by a moving optical tweezer with velocity $v$. (b) Transition from a Fermi singularity at $v \ll v_F$ to a polaron quasiparticle at $v \gg v_F$, shown by spectral function $A(\omega;v)$ with increasing $v$ at the interaction parameter $g=v_F$, where $v_F$ is the Fermi velocity. White curves mark the upper and lower half-maximum frequencies of the spectral peak. \label{Fig:Fig1}}
\end{figure}

In this Letter, we explore an alternative mechanism for polaron formation based on rapid, externally prescribed motion of heavy impurities. This idea is inspired by recent advances in optical-tweezer technology, which enabled the controlled transport of individual atoms along specified trajectories. As schematically illustrated in Fig. \ref{Fig:Fig1}(a), dynamically translated optical tweezers and optical conveyor belts provide complementary means of controlling atomic motion \cite{SimonMatthies2024PRA,AhnHwang2025Optica,Lu2026Optica}, suggesting a feasible route towards experiments in which a single impurity is transported through an ultracold Fermi gas at a controlled velocity. Such a setup would provide a way to probe the many-body response to a moving scattering potential. In particular, the infinite-mass limit describes an impurity whose motion is entirely prescribed by external control: though momentum is transferred to the Fermi gas, the resulting change in the impurity velocity vanishes as its mass tends to infinity.

Here, we investigate the spectroscopic response of a one-dimensional ideal Fermi gas to an infinitely heavy impurity moving at constant velocity. Using an exact functional-determinant approach \cite{Demler2012PRX, Schmidt2018Review, Jia2022PRL,Jia2022PRA,Jia2022Review}, we uncover a strongly nonmonotonic many-body response to the prescribed impurity motion. As reported in Fig. \ref{Fig:Fig1}(b), coherence is maximally suppressed around the Fermi velocity but re-emerges at high velocity as a long-lived polaron resonance. We analytically determine the resonance energy and linewidth in the high-velocity limit and develop a microscopic wave-packet picture that elucidates the dynamical suppression of the OC and the polaron coherence restoration by rapid impurity motion, even in the absence of impurity recoil.

We consider a one-dimensional ideal Fermi gas of $N$ fermions of mass $m$ in a box of length $L$, coupled to an infinitely heavy impurity moving at velocity $v$. The impurity has two internal states, $\left| \uparrow \right \rangle$ and $\left|\downarrow \right>$, which are respectively interacting and non-interacting with the Fermi gas. Setting $\hbar=1$, the Hamiltonian in the laboratory frame reads $\widetilde H(t)= | \uparrow \rangle \langle \uparrow | \widetilde H_\uparrow(t)
+|\downarrow\rangle\langle\downarrow|\widetilde H_\downarrow,$
where
\begin{equation}
\begin{aligned}
\widetilde H_\downarrow
&=\sum_k\frac{k^2}{2m}c_k^\dagger c_k,\\
\widetilde H_\uparrow(t)
&=\widetilde H_\downarrow+
g\int dx\,\delta(x-vt)c_x^\dagger c_x .
\end{aligned} 
\end{equation}
Here, $g>0$ denotes the repulsive contact interaction strength, and the field operators in position and momentum space are related by the usual Fourier transformation. We work in the frame co-moving with the impurity by applying the unitary transformation $S(t)=e^{-ivtP}$, where $P=\sum_k k c_k^\dagger c_k$ is the total momentum of the Fermi gas. The transformed Hamiltonian is $H=S^\dagger\widetilde H S-iS^\dagger\partial_tS$, giving $H=|\uparrow\rangle\langle\uparrow|H_\uparrow+ \downarrow\rangle\langle\downarrow|H_\downarrow,$ with
\begin{equation}
\begin{aligned}
H_\downarrow&=\sum_k
\left(\frac{k^2}{2m}-kv\right)c_k^\dagger c_k,\\
H_\uparrow&=H_\downarrow+
g\int dx\,\delta(x)c_x^\dagger c_x.
\end{aligned} \label{eq:H_v}
\end{equation}
The moving impurity thus becomes a static scattering potential in the co-moving frame, while the fermion dispersion acquires the Doppler shift $-kv$.

We characterize the spectroscopic response of the impurity by the time-dependent overlap \cite{Demler2012PRX,Schmidt2018Review}
\begin{equation}
  S(t; v)=\operatorname{Tr}_{F}
\left[\rho_F e^{iH_\downarrow t}
e^{-iH_\uparrow t}\right],
\end{equation}
where $\rho_F$ is the initial density operator of the Fermi gas. The corresponding spectral response is
\begin{equation}
A(\omega;v)=\frac{1}{\pi}\operatorname{Re}
\int_0^\infty dt\,e^{i\omega t}S(t; v),
\end{equation}
where $\omega$ is measured relative to the bare internal-state transition frequency. The prefactor $1/\pi$ ensures the normalisation $\int_{-\infty}^\infty d\omega A(\omega;v) = 1$. The spectral function $A(\omega;v)$ characterizes the energy-resolved transitions associated with coupling the impurity to the interacting state. These response functions can be probed through complementary spectroscopic techniques: Ramsey spectroscopy accesses the time-domain coherence; while its Fourier transform, such as injection radio-frequency (RF) spectroscopy, probes the frequency-dependent transfer rate in the linear-response regime.

Since the Hamiltonian in Eq. (\ref{eq:H_v}) contains only fermionic, bilinear many-body operators, the many-body overlap can be evaluated exactly using the functional-determinant approach (FDA) \cite{Schmidt2018Review,Jia2022Review}. In terms of the single-particle Hamiltonians $h_\uparrow$ and $h_\downarrow$, and the initial single-particle occupation operator $\hat n$, the Ramsey response is
\begin{equation}
S(t; v)=\det\left[
1-\hat n+
e^{ih_\downarrow t}e^{-ih_\uparrow t}\hat n
\right]. \label{Eq:FDA}
\end{equation}
This determinant incorporates arbitrarily many particle-hole excitations and requires no truncation of the many-body wave function. The corresponding single-particle operators are
\begin{equation}
h_\downarrow=\frac{p^2}{2m}-vp,\ \ h_\uparrow=h_\downarrow+g\delta(x),\label{Eq:hv}
\end{equation}
with $p=-i\partial_x$. We impose periodic boundary conditions on the original finite system and take the thermodynamic limit $N,L\rightarrow\infty$, while keeping the density $n=N/L$ fixed. At zero temperature, $\hat n$ projects onto the occupied momentum states $|k|\leq k_F$, with $k_F=\pi n$ for a single-component Fermi gas. The corresponding Fermi energy and Fermi velocity are given by $E_F=k_F^2/(2m)$ and $v_F=k_F/m$, respectively. Consequently, the ratio of the characteristic interaction energy to the Fermi energy can be expressed as $gn/E_F=(2/\pi)g/v_F$. Within this framework, Eq. (\ref{Eq:FDA}) allows an exact evaluation of the Ramsey response at arbitrary impurity velocity, including all particle-hole excitations.

\begin{figure}[t]
  \centering
  \includegraphics[width=\columnwidth]{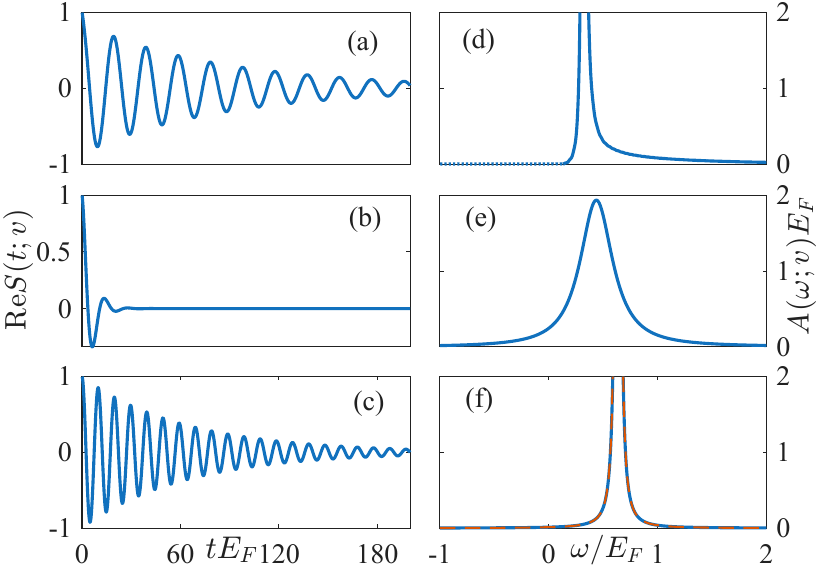}
  \caption{(a)–(c) Real part of the Ramsey response, $\mathrm{Re},S(t;v)$, for impurity velocities $v/v_F=0.05$, $1$, and $20$, respectively. (d)–(f) Corresponding dimensionless spectral functions $E_F A(\omega;v)$. The interaction strength is $g/v_F=1$. In (f), the dashed curve shows a Lorentzian reconstructed from the long-time fit to the Ramsey response, in excellent agreement with the numerical spectrum. \label{Fig:Fig2}}
\end{figure}

We first examine the Ramsey response and the corresponding spectral function for three representative impurity velocities, as shown in Fig. \ref{Fig:Fig2}. We set $g=k_F/m\equiv v_F$ and consider $v/v_F=0.05$, $1$, and $20$. The time-domain response exhibits damped oscillations at low and high velocities, whereas near $v=v_F$ the oscillations decay much more rapidly [Figs. \ref{Fig:Fig2}(a)–\ref{Fig:Fig2}(c)]. At long times, the response is well described by
\begin{equation}
S(t;v)\simeq D(v)e^{-iE_p(v)t-E_\Gamma(v)t},
\end{equation}
where $E_p(v)$ determines the oscillation frequency, $E_\Gamma(v)$ is the decay rate, and $Z(v) \equiv |D(v)|$ is the amplitude of the exponential contribution. This form corresponds to a Lorentzian spectral resonance with energy $E_p(v)$, half-width at half-maximum $E_\Gamma(v)$, and spectral weight $Z(v)$, within the normalization convention used here. Details of fitting procedure are given in Appendix \ref{Sec:Fit}.

The corresponding spectra exhibit a striking nonmonotonic dependence on impurity velocity [Figs. \ref{Fig:Fig2}(d)–2(f)]. At low velocity, the spectrum displays a sharp but strongly asymmetric feature reminiscent of the Fermi-edge singularity associated with a static scattering potential. Near $v=v_F$, it broadens substantially, consistent with the rapid decay of the Ramsey response. In contrast, at high velocity a narrow, approximately Lorentzian resonance emerges. As shown in Fig. \ref{Fig:Fig2}(f), a Lorentzian reconstructed from the long-time Ramsey response agrees closely with the full numerical spectrum. We identify this feature as a long-lived polaron resonance, characterised by $E_p(v)$, the coherence decay rate $E_\Gamma(v)$ and spectral weight $Z(v)$. Its emergence is particularly notable because the impurity is infinitely heavy and undergoes no recoil. The white curves in Fig. \ref{Fig:Fig1}(b) trace the half-maximum frequencies and hence the spectral full width at half maximum (FWHM), which approaches $2E_\Gamma(v)$ at high velocity, consistent with the Lorentzian line shape inferred from the Ramsey response.

\begin{figure*}[t]
  \centering
  \includegraphics[width=\textwidth]{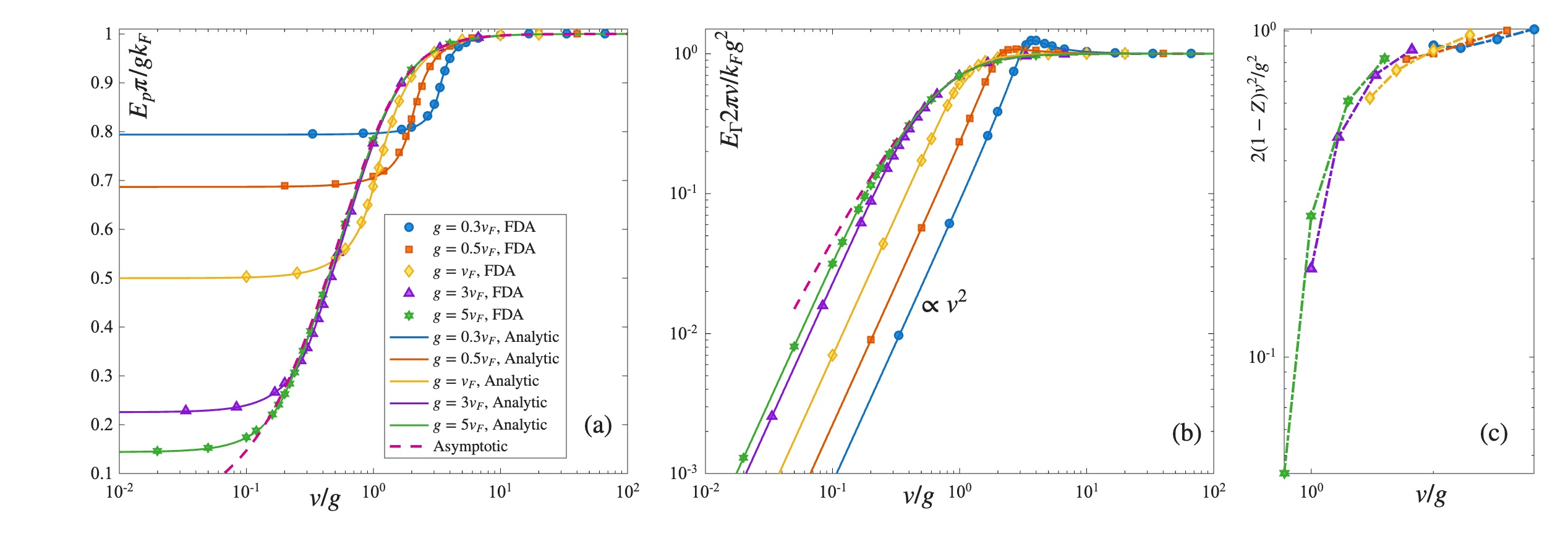}
  \caption{Polaron energy, decay rate, and residue as functions of the impurity velocity $v$ for different interaction strengths $g$. (a) Normalized polaron energy $E_p\pi/(gk_F)$, (b) normalized decay rate $2\pi v E_\Gamma/(k_Fg^2)$, and (c) normalized deviation of the residue from unity, $2(1-Z)v^2/g^2$, plotted against $v/g$. Symbols represent numerically exact results extracted from the functional determinant approach (FDA), solid curves show the analytical results, and dashed curves show the corresponding asymptotic expressions. The dash-dotted curves in (c) are guides to the eye. Different colors and symbols correspond to $g=0.3v_F$, $0.5v_F$, $v_F$, $3v_F$, and $5v_F$, as indicated in the legend of (a). \label{Fig:Fig3}}
\end{figure*}

We now characterise the velocity dependence of the polaron energy $E_p$, decay rate $E_\Gamma$, and spectral weight $Z$, extracted from the long-time Ramsey response. Figure \ref{Fig:Fig3} shows these quantities for several interaction strengths. The polaron energy approaches finite limits in both the low- and high-velocity regimes, whereas the decay rate exhibits distinct asymptotic behaviours: $E_\Gamma\propto v$ at low velocity and $E_\Gamma\propto 1/v$ at high velocity. The spectral weight also increases towards unity as the impurity velocity increases, indicating the emergence of a coherent resonance. The scaling in Fig. \ref{Fig:Fig3} suggests a simple analytical description. We now derive $E_p(v)$ and $E_\Gamma(v)$ over the full velocity range using a long-time scattering picture, whose high-velocity limit explains the suppression of decoherence.

The high-velocity asymptotics of $E_p$ and $E_\Gamma$ can be understood most naturally in a further Galilean-boosted frame. The linear momentum term in the single-particle Hamiltonians in Eq. (\ref{Eq:hv}) is eliminated by the transformation $U_v=e^{imvx}$, yielding
\begin{equation}
U_v^\dagger h_\downarrow U_v
=\frac{p^2}{2m}-\frac{mv^2}{2},\ 
U_v^\dagger h_\uparrow U_v
=\frac{p^2}{2m}+g\delta(x)-\frac{mv^2}{2}. \label{eq:h_boost}
\end{equation}
The two Hamiltonians in Eq. (\ref{eq:h_boost}) differ only by the static delta-function potential, while the initial Fermi sea is shifted to the momentum interval $[-k_F-mv,k_F-mv]$. This observation allows us to derive analytical expressions for the long-time polaron energy $E_p(v)$ and exponential decay rate $E_\Gamma(v)$ over the full velocity range.

We begin by considering the scattering states of the static potential in the boosted frame. At a given positive scattering momentum $k$, there are two incoming channels, corresponding to particles incident from the left and right \cite{Levitov1996JMP}. The scattering matrix is
\begin{equation}
  S_{\rm sc}(k)=
\begin{bmatrix}
t_{\rm sc}(k)&r_{\rm sc}(k)\\
r_{\rm sc}(k)&t_{\rm sc}(k)
\end{bmatrix},
\end{equation}
where
\begin{equation}
  t_{\rm sc}(k)=\frac{k}{k+img},\qquad
r_{\rm sc}(k)=-\frac{img}{k+img}.
\end{equation}
The occupations of the two incoming channels depend on the impurity velocity. For $v<v_F$, both channels are occupied over $0<k<k_F-mv$, whereas only one channel is occupied over $k_F-mv<k<k_F+mv$. For $v>v_F$, the doubly occupied interval disappears, and the singly occupied interval becomes $mv-k_F<k<mv+k_F$.

In the long-time wave-packet description \cite{Levitov1996JMP}, an energy interval $dE$ contains approximately $tdE/(2\pi)$ independent temporal modes during an observation time $t$. Their extensive contribution to the many-body overlap is determined by the energy-resolved scattering matrix (see Appendix \ref{Sec:WavePacket}). When both incoming channels are occupied, the relevant determinant is $\det S_{\rm sc}(k)$; since scattering is unitary, $|\det S_{\rm sc}(k)|=1$, these modes contribute only to the phase. When a single channel is occupied, the determinant reduces to the transmission amplitude $t_{\rm sc}(k)$, whose magnitude is smaller than unity and therefore generates exponential decay. This yields
\begin{equation}
E_p(v)=\frac{F(k_F+mv)+\operatorname{sgn}(k_F-mv)F(|k_F-mv|)}{2\pi m}, \label{eq:Epfull}
\end{equation}
and
\begin{equation}
E_\Gamma(v)=
\frac{G(k_F+mv)-G(|k_F-mv|)}{4\pi m}. \label{eq:EGammafull}
\end{equation}
Here, for $K\geq0$,
\begin{equation}
F(K)=\frac{1}{2}\left[
K^2{\rm atan}\left(\frac{mg}{K}\right)
+mgK-(mg)^2{\rm atan}\left(\frac{K}{mg}\right)
\right],
\end{equation}
and
\begin{equation}
  G(K)=\frac12\left[
K^2\ln\left(1+\frac{(mg)^2}{K^2}\right)
+(mg)^2\ln\left(1+\frac{K^2}{(mg)^2}\right)
\right],
\end{equation}
with $F(0)=G(0)=0$ understood by continuity.

Within the long-time wave-packet approximation, these expressions describe the full velocity dependence of $E_p$ and $E_\Gamma$ and agree remarkably well with the values extracted independently from the exact FDA response. The expressions also reveal the distinct roles of the two scattering-channel configurations. Doubly occupied modes contribute to the phase accumulation, whereas the exponential decay is determined entirely by the singly occupied interval. At $v=v_F$ the doubly occupied interval closes while the singly occupied interval reaches $k=0$, where scattering is strongest. This kinematic crossover explains the maximal decoherence around the Fermi velocity. For $v>v_F$, the singly occupied interval progressively shifts towards larger scattering momenta, where the effective scattering becomes weaker, thereby reducing the decay and ultimately restoring coherence at high velocity.

At high velocity, $v\gg v_F$, Eqs. (\ref{eq:Epfull}) and (\ref{eq:EGammafull}) simplify to
\begin{equation}
  E_p(v)\simeq
\frac{k_Fv}{\pi}\arctan\left(\frac{g}{v}\right),
\end{equation}
and
\begin{equation}
  E_\Gamma(v)\simeq
\frac{k_Fv}{2\pi}
\ln\left[1+\left(\frac{g}{v}\right)^2\right].
\end{equation}
These are the asymptotic expressions plotted as dashed curves in Figs. \ref{Fig:Fig3}(a) and \ref{Fig:Fig3}(b), while the solid curves show the full analytical results. In the further limit $v\gg g$, the expressions reduce to
\begin{equation}
E_p(v)\simeq gn,\qquad
E_\Gamma(v)\simeq\frac{k_Fg^2}{2\pi v}.
\end{equation}
Thus, the polaron energy approaches a finite mean-field value, while the decay rate vanishes inversely with velocity. This behaviour follows from the different contributions of the transmission amplitude: The phase of the transmission amplitude produces the energy shift at first order in $g/v$, whereas its departure from unit magnitude enters only at second order. Consequently, $E_p \rightarrow gn$ while $E_\Gamma \propto g^2/v$, explaining how rapid motion restores a long-lived resonance.

The spectral weight provides a complementary measure of coherence restoration. As shown in Fig. \ref{Fig:Fig3}(c), it approaches unity at high velocity, with $1-Z\propto g^2/v^2$. Together with the vanishing decay rate, this behaviour indicates that rapid prescribed motion suppresses the many-body decoherence associated with the orthogonality catastrophe. Importantly, this mechanism does not rely on impurity recoil or on opening a gap in the surrounding medium.

The predicted regime is accessible at experimentally relevant velocities. For representative parameters of a one-dimensional $^6$Li gas, we take $k_F\simeq6.9\mu{\rm m}^{-1}$, consistent with experimentally realized densities \cite{HuletYang2018PRL}, corresponding to $v_F\simeq0.073{\rm m/s}$, so that $20v_F\simeq1.5{\rm m/s}$. Recent optical-tweezer experiments have demonstrated atom transport at velocities exceeding $4.2 {\rm m/s}$ \cite{Lu2026Optica}. For $g=v_F$, our high-velocity expression gives $E_\Gamma/h\simeq0.63{\rm kHz}$ and $E_\Gamma/E_p\simeq2.5\%$ at $v=20v_F$. These scales place the predicted narrow resonance in a regime amenable to time-domain Ramsey measurements or frequency-resolved impurity spectroscopy.

In summary, we have investigated the spectroscopic response of an infinitely heavy impurity moving at constant velocity through a one-dimensional ideal Fermi gas. The exact FDA captures the evolution from an asymmetric low-velocity spectrum to a broadened response near the Fermi velocity and a narrow resonance at high velocity. Analytical expressions for the polaron energy and decay rate reveal the full velocity dependence and recover the previously identified high-velocity asymptotics. The results establish controlled impurity motion as a mechanism for suppressing orthogonality-catastrophe-induced decoherence and restoring a long-lived polaron resonance without impurity recoil. The underlying high-velocity mechanism might also extend to higher dimensions, although the corresponding phase-space structure and quantitative coherence restoration remain to be explored.

\begin{acknowledgments}
  This research was supported by the Australian Research Council's (ARC) Discovery Program, Grants Nos. FT230100229 (J.W.), DP240100248 (X.-J.L.), and DP240101590 (H.H.).
\end{acknowledgments}

\bibliography{VPolaron}

\appendix
\section{Fitting procedure \label{Sec:Fit}}
Numerically, $E_p$, $E_\Gamma$, and $Z$ are extracted from the asymptotic form $S(t;v)\approx D e^{-i E_p t - E_\Gamma t}$. Unless otherwise specified, we use the fitting window $40 \leq tE_F \leq160$. We first determine $Z$ and $E_\Gamma$ from a fit of the magnitude, $|S(t;v)| = Z e^{-E_\Gamma t}$. Keeping these parameters fixed, we then fit the normalised complex response $S(t;v)/\left(Ze^{-E_\Gamma t}\right)$ to $e^{-i E_p t + i\phi_0}$, from which $E_p$ and the phase $\phi_0$ are obtained, with $D = Ze^{i\phi_0}$. We have checked the robustness of this procedure by varying both boundaries of the fitting interval and the system size. To minimise finite-size effects associated with the moving impurity, we additionally choose the system size such that $L > v t_f$ throughout the calculations.

\section{Full velocity dependence from the wave-packet representation \label{Sec:WavePacket}} 

We derive analytical expressions for the polaron energy $E_p(v)$ and decay rate $E_\Gamma(v)$ using a wave-packet representation of scattering states, following the approach of Levitov \textit{et al.}~\cite{Levitov1996JMP}. The derivation evaluates the extensive, linear-in-time contribution to the logarithm of the Ramsey response in the long-time limit.

After applying the Galilean boost $U_v=e^{imvx}$, the single-particle Hamiltonians become
\begin{equation}
h'_\downarrow=\frac{p^2}{2m}-\frac{mv^2}{2},
\qquad
h'_\uparrow=\frac{p^2}{2m}+g\delta(x)-\frac{mv^2}{2}.
\end{equation}
The common energy shift cancels in the relative evolution entering the Ramsey response. The initial Fermi sea occupies the shifted momentum interval
\begin{equation}
-k_F-mv\leq \tilde k\leq k_F-mv.
\end{equation}
For each positive scattering momentum $k = |\tilde k|$, we introduce two incoming channels, $L$ and $R$, corresponding to particles incident on the potential from the left and right, respectively. Thus, for $v<v_F$, both channels are occupied over $0<k<k_F-mv$, whereas only one channel is occupied over $k_F-mv<k<k_F+mv$. For $v>v_F$, the doubly occupied interval disappears, leaving the singly occupied interval $mv-k_F<k<mv+k_F$.

For the static delta-function potential in the boosted frame, the scattering states have energy $E_k=k^2/(2m)$, where the common energy shift has been omitted. Their transmission and reflection amplitudes are
\begin{equation}
t_{\rm sc}(k)=\frac{k}{k+img},
\qquad
r_{\rm sc}(k)=-\frac{img}{k+img}.
\end{equation}
The corresponding scattering matrix, in the channel ordering $(L,R)$, is
\begin{equation}
S_{\rm sc}(k)=
\begin{bmatrix}
t_{\rm sc}(k)&r_{\rm sc}(k)\\
r_{\rm sc}(k)&t_{\rm sc}(k)
\end{bmatrix}.
\end{equation}
For real $k$ and $g$, this matrix is unitary.

Following the temporal-mode construction of Ref.~\cite{Levitov1996JMP}, we represent the incoming states by wave packets labelled by their arrival times $\tau$ at the scattering potential. In the idealized limit, the packet envelopes are delta functions propagating at the group velocity $v_g=k/m$. Before reaching the potential, the packets evolve as freely propagating states; after scattering, each incoming packet separates into transmitted and reflected components with amplitudes $t_{\rm sc}(k)$ and $r_{\rm sc}(k)$.

More precisely, the delta-function packets should be understood as an ideal temporal-mode representation rather than as ordinary wave packets that are simultaneously localized in time and energy. They can be regularized using packets with a finite energy width \cite{Schmidt2018Review}. In the long-time limit, the energy intervals can be made sufficiently narrow that the scattering amplitudes are approximately constant within each interval. The dispersion can then be linearized locally, so that packets associated with different arrival times become orthogonal to leading order. Under these assumptions, the extensive contribution to the logarithm of the FDA determinant can be evaluated by treating the temporal modes independently.

The exact single-particle operator entering the FDA is
\begin{equation}
\hat B(t)=
1-\hat n+
e^{ih'_\downarrow t}e^{-ih'_\uparrow t}\hat n,
\end{equation}
where $\hat n$ is the initial occupation operator. In the long-time wave-packet approximation, the leading extensive contribution to $\ln\det\hat B(t)$ is determined by the channel-space overlap matrix
\begin{equation}
\mathcal M(k)=
I-\mathrm n(k)+S_{\rm sc}(k)\mathrm n(k),
\end{equation}
with
\begin{equation}
\mathrm n(k)=
\begin{pmatrix}
n_L(k)&0\\
0&n_R(k)
\end{pmatrix}.
\end{equation}

In the long-time limit, an energy interval $dE_k$ contains approximately $t dE_k/(2\pi)= kdk/(2\pi m)$, independent temporal modes. The contribution to $\ln S (t;v)$ that is extensive in time is therefore
\begin{equation}
\left.\ln S(t ; v)\right|_{O(t)}=\frac{t}{2 \pi m} \int_0^{\infty} k d k \ln \operatorname{det} {\mathcal M}(k).
\label{eq:lnS_wavepacket}
\end{equation}
Subleading contributions, including possible algebraic factors associated with the orthogonality catastrophe, are not determined by this temporal-mode counting.

The determinant depends on the occupation of the two incoming channels. When neither channel is occupied, $\det\mathcal M(k)=1$, so the corresponding modes make no contribution. When only one channel is occupied, the determinant reduces to the transmission amplitude,
\begin{equation}
\det\mathcal M(k)=t_{\rm sc}(k).
\end{equation}
When both channels are occupied, it reduces to the determinant of the scattering matrix,
\begin{equation}
\det\mathcal M(k)
=\det S_{\rm sc}(k)
=t_{\rm sc}^2(k)-r_{\rm sc}^2(k)
=\frac{k-img}{k+img}.
\end{equation}
Because the scattering matrix is unitary, $|\det S_{\rm sc}(k)|=1$. Thus, the doubly occupied interval contributes to the phase of the many-body overlap but not to its extensive exponential decay. By contrast, the singly occupied interval contributes to both phase accumulation and decay because $|t_{\rm sc}(k)|<1$ for nonzero $g$.

Comparing Eq.~(\ref{eq:lnS_wavepacket}) with the long-time form
\begin{equation}
S(t;v)\simeq D(v)e^{-iE_p(v)t-E_\Gamma(v)t},
\end{equation}
we obtain the decay rate
\begin{equation}
E_\Gamma(v)=
-\frac{1}{2\pi m}
\int_{\rm single}kdk\,
\ln|t_{\rm sc}(k)|.
\end{equation}
For the delta-function potential,
\begin{equation}
|t_{\rm sc}(k)|^2
=\frac{k^2}{k^2+(mg)^2},
\end{equation}
and hence
\begin{equation}
E_\Gamma(v)=
\frac{1}{4\pi m}
\int_{|k_F-mv|}^{k_F+mv}
dk\,k\ln\left[1+\frac{(mg)^2}{k^2}\right].
\label{eq:Gamma_integral}
\end{equation}

The polaron energy follows from the phase of the determinant. For repulsive interactions, $g>0$, the transmission amplitude has phase
\begin{equation}
\arg t_{\rm sc}(k)
=-\arctan\left(\frac{mg}{k}\right),
\end{equation}
while the phase of the determinant for two occupied channels is
\begin{equation}
\arg\det S_{\rm sc}(k)
=-2\arctan\left(\frac{mg}{k}\right),
\end{equation}
with the phase branch chosen continuously from $g=0$. It follows that
\begin{widetext}
\begin{equation}
E_p(v)=
\frac{1}{2\pi m}
\begin{cases}
\displaystyle
\int_0^{k_F+mv}dk\,k
\arctan\left(\frac{mg}{k}\right)
+
\int_0^{k_F-mv}dk\,k
\arctan\left(\frac{mg}{k}\right),
& v<v_F,\\
\displaystyle
\int_{mv-k_F}^{mv+k_F}dk\,k
\arctan\left(\frac{mg}{k}\right),
& v>v_F.
\end{cases}
\label{eq:Ep_integral}
\end{equation}
\end{widetext}

To express these results in closed form, we introduce the functions
\begin{equation}
F(K)=
\int_0^K dk\,k\arctan\left(\frac{mg}{k}\right),
\end{equation}
and
\begin{equation}
G(K)=
\int_0^K dk\,k\ln\left[1+\frac{(mg)^2}{k^2}\right],
\end{equation}
for $K\geq0$. Evaluating the integrals gives
\begin{equation}
F(K)=\frac12\left[
K^2\arctan\left(\frac{mg}{K}\right)
+mgK-(mg)^2\arctan\left(\frac{K}{mg}\right)
\right],
\end{equation}
and
\begin{equation}
G(K)=\frac12\left[
K^2\ln\left(1+\frac{(mg)^2}{K^2}\right)
+(mg)^2\ln\left(1+\frac{K^2}{(mg)^2}\right)
\right],
\end{equation}
where $F(0)=G(0)=0$ by continuity. Equations~(\ref{eq:Gamma_integral}) and (\ref{eq:Ep_integral}) then give
\begin{equation}
E_p(v)=
\frac{
F(k_F+mv)+
\operatorname{sgn}(k_F-mv)F(|k_F-mv|)
}{2\pi m},
\label{eq:Ep_full}
\end{equation}
and
\begin{equation}
E_\Gamma(v)=
\frac{
G(k_F+mv)-G(|k_F-mv|)
}{4\pi m}.
\label{eq:Gamma_full}
\end{equation}
These expressions describe the full velocity dependence within the long-time wave-packet approximation.
\end{document}